\documentclass[usegraphicx]{emulateapj}
\usepackage{natbib}
\usepackage{color}
\usepackage{xspace}
\usepackage[]{units}
\usepackage{verbatim}
\shorttitle{Dark Matter Annihilation Over Cosmic Time}
\shortauthors{Mack}

\begin{document}
\bibliographystyle{apj_short}

\submitted{Submitted to MNRAS}

\newenvironment{inlinefigure}{
\def\@captype{figure}
\noindent\begin{minipage}{0.999\linewidth}\begin{center}
{\end{center}\end{minipage}\smallskip}}

\newcommand{\pd}{\partial}
\newcommand{\rms}{r_{\rm ms}}
\newcommand{\bsym}[1]{\mbox{\boldmath$#1$}}
\newcommand{\be}{\begin{equation}}
\newcommand{\ee}{\end{equation}}
\newcommand{\ud}{\mathrm{d}}
\newcommand{\msun}{M$_\odot$}

\newcommand{\red}{\color{red}}
\newcommand{\blue}{\color{blue}}
\newcommand{\green}{\color{PineGreen}}

\title{Known Unknowns of Dark Matter Annihilation Over Cosmic Time}

\author{Katherine J. Mack$^{*1,2,3}$}
\altaffiltext{1}{School of Physics (David Caro Building), University of Melbourne, Victoria 3010, Australia}
\altaffiltext{2}{ARC Centre of Excellence for All-sky Astrophysics (CAASTRO)}
\altaffiltext{3}{ARC Centre of Excellence for Particle Physics at Terascale (CoEPP)}
\email[*]{kmack@unimelb.edu.au}

\begin{abstract}
Dark matter self-annihilation holds promise as one of the most robust mechanisms for the identification of the particle responsible for the Universe's missing mass. In this work, I examine the evolution of the dark matter annihilation power produced by smooth and collapsed structures  over cosmic time, taking into account uncertainties in the structure of dark matter halos. As we search for observational signatures of annihilation, an understanding of this time evolution will help us to best direct our observational efforts, either with local measurements or investigation of the effects of annihilation on the intergalactic medium at high redshift. As I show in this work, there are several key sources of uncertainty in our ability to estimate the dark matter annihilation from collapsed structures, including: the density profile of dark matter halos; the small-scale cut-off in the dark matter halo mass function; the redshift-dependent mass-concentration relation for small halos; and the particle-velocity dependence of the dark matter annihilation process. Varying assumptions about these quantities can result in annihilation power predictions that differ by several orders of magnitude. These uncertainties must be resolved, through a combination of observation and modeling, before robust estimations of the cosmological annihilation signal can be made.

\end{abstract}

\section{Introduction}\label{sec:intro}

While it is widely accepted that dark matter makes up the bulk of the mass in the Universe, the nature of the dark matter particle is still unknown. Currently, dark matter identification efforts are faced with a large set of inconclusive or contradictory results in experiments for indirect \citep{Abdo2009,Adriani2009,Aguilar2013} and direct detection \citep{Aalseth2011,Aalseth2013,Ahmed2009,Ahmed2010,Ahmed2011,Angle2008,Angle2011,Angloher2012,Armengaud2012,Bernabei2010,Kelso2012,LUXCollaboration2013,Savage2009}. This confusing state of affairs highlights the importance of seeking alternative probes of dark matter's properties. In this work, I discuss the uncertainties and prospects of one promising avenue for exploration of dark matter: modelling the consequences of dark matter annihilation on the local baryonic medium in which dark matter halos reside \citep{Padmanabhan2005,Furlanetto2006,Valdes2012}. The form and ultimate impact of the energy deposition when annihilation occurs in a baryonic medium depend strongly on density, temperature, ionization state, and redshift. However, it is nonetheless instructive to ask the question of how much {\it total} annihilation power is being produced by dark matter halos at a given redshift (per unit volume). Knowing this evolution can be useful in determining if there is a ``sweet spot'' in which to look for signs of local energy injection, where the total annihilation power is higher than at other times in the evolution of structure. Also, calculating this evolution is a first step to understanding the overall impact of dark matter annihilation on baryonic structures and the evolution of stars and galaxies.

As I show in this paper, there are several major uncertainties that must be addressed before this question can be answered. Specifically, the form and abundance of dark matter halos, especially at the lowest masses, must be known in order for the annihilation power to be determined. Since the structure of dark matter halos is difficult to observe directly or to simulate at mass scales below those of galaxy clusters, our knowledge of low-mass structures is generally an extrapolation from the properties of halos at higher masses \citep{Coe2008}. The choices made in these extrapolations over several orders of magnitude strongly affect predictions of the annihilation power produced, making it necessary to more carefully examine these assumptions before we can make robust predictions to apply to future observations. [For an alternative approach to parameterizing these model dependencies, see \cite{Serpico2012}.]

This paper is structured as follows: In Section \ref{sec:dm}, I review basic properties of dark matter annihilation and the expected signal for smoothly distributed and collapsed structures. In Section \ref{sec:model}, I describe the dark matter halo modelling I used in this work, with Sections \ref{subsec:massfunction}, \ref{subsec:densityprofile}, \ref{subsec:minmass} and \ref{subsec:concentration} describing my modelling for the halo mass function, density profile, minimum mass cut-off, and mass-concentration relations, respectively. Section \ref{sec:results} presents the results of the modelling for different choices of dark matter halo models, with Sections \ref{subsec:massfunctionres}-\ref{subsec:concentrationres} following the sequence of Sections \ref{subsec:massfunction}-\ref{subsec:concentration}. Secton \ref{subsec:range} discusses the full range of predictions for models considered here. In Section \ref{sec:disc}, I discuss the implications of these results, with Section \ref{subsec:haloprops} summarizing the effects of different halo models and Section \ref{subsec:velocity} discussing how the consideration of a velocity-dependent annihilation rate would also affect these predictions.

Throughout the work, I have assumed cosmological parameters derived from the first Planck cosmology release \citep{PlanckCollaboration2013}, using the derived cosmological parameter set ``Planck+WP+highL+BAO,'' specifically: $h= 0.6777$, $\Omega_{b,0} h^2=0.022161$, $\Omega_{c,0} h^2 = 0.11889$, $\sigma_8=0.8288$.

\section{Dark matter annihilation}\label{sec:dm}

The annihilation rate per unit volume of weakly interacting massive particle (WIMP) dark matter is proportional to the square of the local dark matter density, as annihilation is a two-particle process. Following \cite{Cirelli2009}, I separate the dark matter annihilation rate density $A(z)$ into two components, a ``smooth'' component $A_{sm}(z)$, accounting for annihilation in uncollapsed regions, and a ``structure'' component $A_{\rm{struct}}(z)$, accounting for annihilation that occurs within collapsed halos:
\begin{eqnarray}
A_{\rm{sm}}(z) & = & \frac{\langle \sigma v \rangle}{2 m_{\chi}^2} \rho_{\rm{DM},0}^2 (1+z)^6  \label{eq:sm} \\
A_{\rm{struct}}(z) & = & \frac{\langle \sigma v \rangle}{2 m_{\chi}^2} \int_{M_{\rm{min}}}^{M_{\rm{max}}} \left[ \frac{\ud n}{\ud M}(z,M) (1+z)^3 \right. \label{eq:struct} \nonumber \\ 
 &  & \left. \int_{V_{\rm{vir}}} \rho_{\rm{DM}}^2(r, z, M) \ud V \right] \ud M
\end{eqnarray}
where $\langle \sigma v \rangle$ is the thermally averaged annihilation cross section, $m_\chi$ is the mass of the dark matter particle, $\rho_{\rm{DM},0}$ is the mean dark matter density at redshift 0, $\frac{\ud n}{\ud M}(z,M)$ is the comoving number density of halos of mass $M$ [in equation (\ref{eq:struct}), the $(1+z)^3$ factor converts this to physical density], and $\rho_{\rm{DM}}(r, z, M)$ represents the dark matter halo radial density profile for a halo of mass $M$ at redshift $z$. The limits of integration $M_{\rm{min}}$ and $M_{\rm{max}}$ over the mass function are the minimum and maximum mass of halos included in the calculation, and the volume integral is over $V_{\rm{vir}}$, the volume out to the virial radius.

To a first approximation, the two components can be thought of as additive 
\be A(z) \approx A_{sm}(z) + A_{\rm{struct}}(z),
\ee
though because the ``smooth'' component relies on the globally averaged dark matter density, it also includes the dark matter that is contained in halos. Therefore simply adding the components both double-counts the dark matter in halos and overestimates the mean density of dark matter in voids. However, in most regimes, the two components are not of similar magnitudes, so this approximation is sufficient.

At redshift $z$, the volume-averaged power produced from dark matter annihilation events can be calculated as
\be
P(z) = 2 m_\chi c^2 A(z) .
\ee
For ease of comparison with other studies \citep[e.g.][]{Padmanabhan2005,Furlanetto2006}, it is useful to state this as energy output per second per hydrogen nucleus: $P_H(z) = P(z)/n_H(z)$ . To a good approximation, the number density of hydrogen nuclei at redshift $z$ can be written as $n_H(z) = \Omega_{b,0} \rho_{\textrm{crit},0} (1-Y_p) (1+z)^3 / m_H$ where $\Omega_{b,0}$ is the present-day baryon density parameter, $\rho_{\textrm{crit},0}$ is the present-day critical density, $Y_p=0.2477$ is the primordial mass fraction of helium \citep{PlanckCollaboration2013}, and $m_H$ is the mass of the proton. Note that the main results here are stated in terms of $h^4$ due to the scaling of masses and volumes in the halo mass functions used here.

\section{Halo model}\label{sec:model}

In order to calculate the annihilation rate of dark matter in collapsed structures using equation \ref{eq:struct}, we must have models for:
\begin{itemize}
\item Dark matter particle properties $\langle \sigma v \rangle$ and $m_{\chi}$
\item Halo mass function $\frac{\ud n}{\ud M}(z,M)$
\item Minimum and maximum mass cut-offs $M_{\rm{min}}$ and $M_{\rm{max}}$
\item Halo density profile $\rho_{\rm{DM}}^2(r, z, M)$
\item Mass-concentration relation $c(z,M)$
\end{itemize}
As I will discuss in the sections below, there are large and important uncertainties in each of these properties that make an estimate of the annihilation output of dark matter from collapsed structures difficult. My aim in this work is to demonstrate the magnitude of these uncertainties using some plausible examples and boundary cases and show how they affect estimates of the dark matter annihilation power.

\subsection{Halo mass function}\label{subsec:massfunction}

In this work, I use the mass functions generated by the ``genmf.f'' mass-function code developed by \cite{Reed2007}\footnote{The code is publicly available online at http://icc.dur.ac.uk/Research/PublicDownloads/genmf\_readme.html}. For the purpose of showing a range of possible outcomes, I consider three different mass functions:
\begin{itemize}
\item \cite{Reed2007}
\item \cite{Press1974} (``Press-Schechter'')
\item \cite{Sheth1999} (``Sheth-Tormen'')
\end{itemize}
These mass functions are chosen because they have been shown to match simulations and observations to a good level of accuracy for a wide range of scales. In all cases, however, the mass function is extrapolated far beyond the range at which we have the means to test or consistently simulate it. As I will discuss below, the lowest-mass halos have a strong effect on the dark matter annihilation signal, and with all three mass functions it has been necessary to assume a simple power-law extrapolation to model masses below about $10^5$ M$_\odot$. 

\subsection{Halo density profile}\label{subsec:densityprofile}

Because dark matter annihilation is a density-squared effect, the form of the halo density profile can affect the energy output even if all models have the same average density (mass $M$ within the virial radius $r_{\rm{vir}}$). In this work, I have considered three density profiles commonly applied to dark matter halos: the Navarro, Frenk \& White (NFW) profile \citep{Navarro1996}, the Einasto profile \citep{Einasto1965,Merritt2006}, and the ``cored'' Burkert profile \citep{Burkert1995}. The functional form of the NFW profile is:
\be
\rho_{\rm{NFW}}(r) = \frac{\rho_s(z,M)}{\frac{r}{r_s(z,M)}\left( 1+ \frac{r}{r_s(z,M)} \right)^2} \label{eq:NFW}
\ee
where $\rho_s(z,M)$ is the density normalization and $r_s(z,M)$ is the scale radius. For the Einasto profile, I use the form
\be
\rho_{\rm{Ein}}(r) = \rho_{E,0}(z, M) \exp{\left( -\frac{2}{\alpha_E} \left[ \left( \frac{r}{r_s(z, M)}\right)^{\alpha_E} - 1 \right]\right)} \label{eq:Ein}
\ee
where $\rho_{E,0}(z,M)$ is the density normalization and $\alpha_E$ is taken to be 0.17. Both these density profiles are normalized such that the total dark matter mass within the virial radius $r_{\rm{vir}}(z,M)$ is $M$. As I show below, the choice between these two halo profiles does not have a strong effect on the overall annihilation signal, as they have been chosen to match observational properties of high-mass dark matter halos, which are generally found to be similar to NFW profiles.

A complication when it comes to choosing a dark matter halo density profile is the possibility that baryonic effects such as supernova feedback can alter the inner density profiles of halos, in some cases creating ``cored'' halos in which the density reaches a plateau in the inner regions \citep{Pontzen2012}. The fact that annihilation power depends on the square of the dark matter density suggests that the shape of the inner profile can strongly influence the overall annihilation output of halos. Other work by \cite{Duffy2010} suggests that in some cases, baryonic physics can increase concentrations in dark matter halos. A better understanding of the role of baryonic feedback in shaping dark matter halos will be necessary for the reliable prediction of the annihilation power over a wide range of masses.

To show how a cored profile could reduce the annihilation signal, I consider the Burkert profile \citep{Burkert1995} as an example:
\be
\rho_{\rm{Burk}}(r) = \frac{\rho_{B,0}(z,M)}{\left(1+\frac{r}{r_0}\right) \left(1+\left(\frac{r}{r_0}\right)^2 \right) } \label{eq:Burk}
\ee
where $r_0=r_{\rm{vir}}/3.4$ and $\rho_{B,0}$ is the density normalization, also chosen so that the total dark matter mass within the virial radius $r_{\rm{vir}}(z,M)$ is $M$. To account for the possibility that cored profiles result only from baryonic feedback effects, I consider two cases: one in which all halos follow equation (\ref{eq:Burk}), and one in which only halos above 1 solar mass have cores (and the rest follow the NFW profile). As will be seen, the latter case, mimicking cored profiles arising from feedback, has little effect on the total annihilation signal.

Some authors have suggested that the form of the density profile derives fundamentally from the halo's accretion history \citep{Ludlow2013a,Angulo2009}. If this is the case, small halos that have not built up by accretion, or that have had very different accretion histories, should not be expected to follow the standard (e.g. NFW) density profiles [but see also \cite{Huss1999}]. Modelling by \cite{Ishiyama2010} suggests microhalo inner profiles can be steeper than NFW. The uncertainty in the applicability of NFW-type halo profiles at low masses or for differing formation histories highlights the need for more detailed simulations of dark matter halos at the smallest masses. As I show below, the annihilation signal from the smallest-mass halos has a large effect on the total power output, and it is also the regime in which our modeling is, at present, least reliable.

\subsection{Minimum mass of dark matter halos}\label{subsec:minmass}

A calculation of dark matter annihilation requires the inclusion of halos down to the lowest mass at which collapsed structures form and persist. This scale is difficult to determine with precision, as it depends upon (1) the primordial power spectrum, (2) the dark matter particle species, (3) the particle mass, and (4) the ability of these microhalos to persist as collapsed objects when subjected to tidal forces during hierarchical merging \citep{Bringmann2009a}. Warm dark matter (WDM) models can alter the minimum mass drastically, but even for cold dark matter, estimates vary widely; \cite{Bringmann2009a} suggests a range from $10^{-11}$ M$_\odot$ to $10^{-4}$ M$_\odot$, based on considering both the free-streaming scale $M_{\rm{fs}}$ and the acoustic oscillation scale $M_{\rm{ao}}$. Looking specifically at neutralino dark matter, \cite{Angulo2009} find a cut-off at $10^{-8}$ M$_\odot$ for particles with masses $m_\chi=100$ GeV, with lower cut-off scales for higher-mass particles. Since my calculation has assumed primordial matter power spectra which extend in principle to arbitrarily low masses, the cut-off due to particle properties of dark matter determines the abundance of the most numerous halos in the Universe. As I will show below, the choice of a low-mass cut-off has a strong effect on the annihilation output, especially at high redshifts. Because the most massive halos are rare, the maximum mass of halos above galaxy scales does not have a significant effect on the total power output.

Previous works [e.g. \cite{Schneider2010}] have explored the effect of microhalo-scale dark matter structures on the annihilation signal in the context of indirect detection and found it to be negligible. It is important to note a key difference between the limits that could be derived from near-term indirect detection experiments and the scenario I explore here. In indirect detection, it is necessary to find a source of a substantial radiation flux in a compact region in order to make the radiation detectable. In this context, small and potentially isolated microhalos produce a radiation flux that is too diffuse to be detected from a distance and removed from the background. Similar microhalos can, however, have a substantial effect on the baryons in their immediate surroundings, and the local energy injection depends strongly on the abundance and size of the dark matter halos. The effects of numerous microhalos at high redshift could contribute to the radiation background and to the heating of local gas in ways that are potentially detectable with observations of the integrated free-electron optical depth to the cosmic microwave background (CMB) \citep{Padmanabhan2005} or with future observations of the 21cm signal of neutral hydrogen \citep{Furlanetto2006,Cirelli2009}.

Because the cut-off scale for cold dark matter is highly dependent on the dark matter model, in this work, I consider the cut-offs $M_{\rm{min}}=10^{-6}$ M$_\odot$ and $M_{\rm{min}}=10^{-9}$ \msun{} as example cases to show the general behavior of varying the cut-off scale.

If dark matter is not cold and collisionless, this would also have a strong effect on the abundance of small-scale halos. Diffusion of dark matter out of the smallest halos (in the case of warm dark matter) or the existence of a pressure (in the case of self-interacting dark matter) would result in a cut-off in the power spectrum at small masses, reducing the annihilation signal and potentially also affecting the build-up of larger halos via accretion. There are some models of WDM that predict self-annihilation \citep{Yuan2012}, so I will also consider how a lower-mass cut-off appropriate to WDM models would affect the annihilation signal. In fitting with current limits on WDM effects from \cite{Viel2013}, I examine cases with $M_{\rm{min}}=10^3$ M$_\odot$ and $M_{\rm{min}}=10^7$ \msun{}.

\subsection{Mass-concentration relation}\label{subsec:concentration}

The concentration parameter determines how concentrated the dark matter is toward the center of a given halo, and is usually defined as $c = r_{\rm{vir}}/r_s$ where $r_s$ is the halo's scale radius. While the mean density of two halos of mass $M$ and radius $r_{\rm{vir}}$ are identical, changes in the concentration parameter can result in changes to the mean {\it squared} density, thus changing the dark matter annihilation radiation output via the $\rho_{\rm{DM}}^2$ term in equation (\ref{eq:struct}). Thus the redshift-dependent mass-concentration relation, $c(z,M)$, has a strong effect on the annihilation power output for a given halo model.

I have considered four example cases of mass-concentration relations. Three relations are chosen to be limiting cases [discussed in the review by \cite{Coe2008}], with very different normalizations and slopes, having been derived from observations of high-mass dark matter halos and simulations of low-mass dark matter halos, respectively. The first mass-concentration relation, from \cite{Comerford2007}, is derived from observations of cluster-mass halos:
\be
c_{CN}(z, M) = \left(\frac{14.5}{1 + z}\right) \left(\frac{M}{1.3 \times 10^{13} h^{-1} \textrm{M}_\odot }\right)^{-0.15}. \label{eq:cCN}
\ee
The second, from \cite{Duffy2008}, is based on numerical simulations down to galaxy mass scales, and has a shallower dependence on both mass and redshift as well as a lower normalization:
\be
c_{D}(z,M) = \left( \frac{7.85}{(1 + z)^{0.71}} \right) \left(\frac{M}{2 \times 10^{12} h^{-1} \textrm{M}_\odot} \right)^{-0.081}. \label{eq:cD}
\ee
Based on the discussion in \cite{Duffy2008}, I also consider a mass-concentration relation based on simulations of halos at redshifts greater than 2 (see Figure 2 in that work), in which I assume an altered version of equation \ref{eq:cD} in which the concentration has no mass dependence:
\be
c_{\rm{flat}}(z,M) = \frac{7.85}{(1 + z)^{0.71}}. \label{eq:cflat}
\ee
This relation, as well as that derived from \cite{Comerford2007} is chosen to demonstrate a limiting case rather than to correspond to a proposed model. It should be noted that the mass-concentration relations of all the cases considered are extrapolated at least 15 orders of magnitude below the masses at which they were derived.

I also consider the mass-concentration relation proposed by \cite{Ludlow2013}, which is based on an examination of halos in the Millennium Simulation \citep{Springel2005,Boylan-Kolchin2009,Angulo2012,Ludlow2013a}. While the relation is modeled self-consistently down to very low masses, it is based only on {\it relaxed} halos in the simulation, and therefore is not necessarily a good approximation at high redshift, where a smaller proportion of halos are relaxed. The derivation of the mass-concentration is discussed in \cite{Ludlow2013}; I use a fitting function in this work. More detailed modelling of the mass-concentration of dark matter microhalos will be necessary to make rigorous predictions of the annihilation output of dark matter, which is strongly influenced by the smallest halos.

\begin{figure}[htbp]
\begin{center}
\resizebox{\columnwidth}{!}{\includegraphics{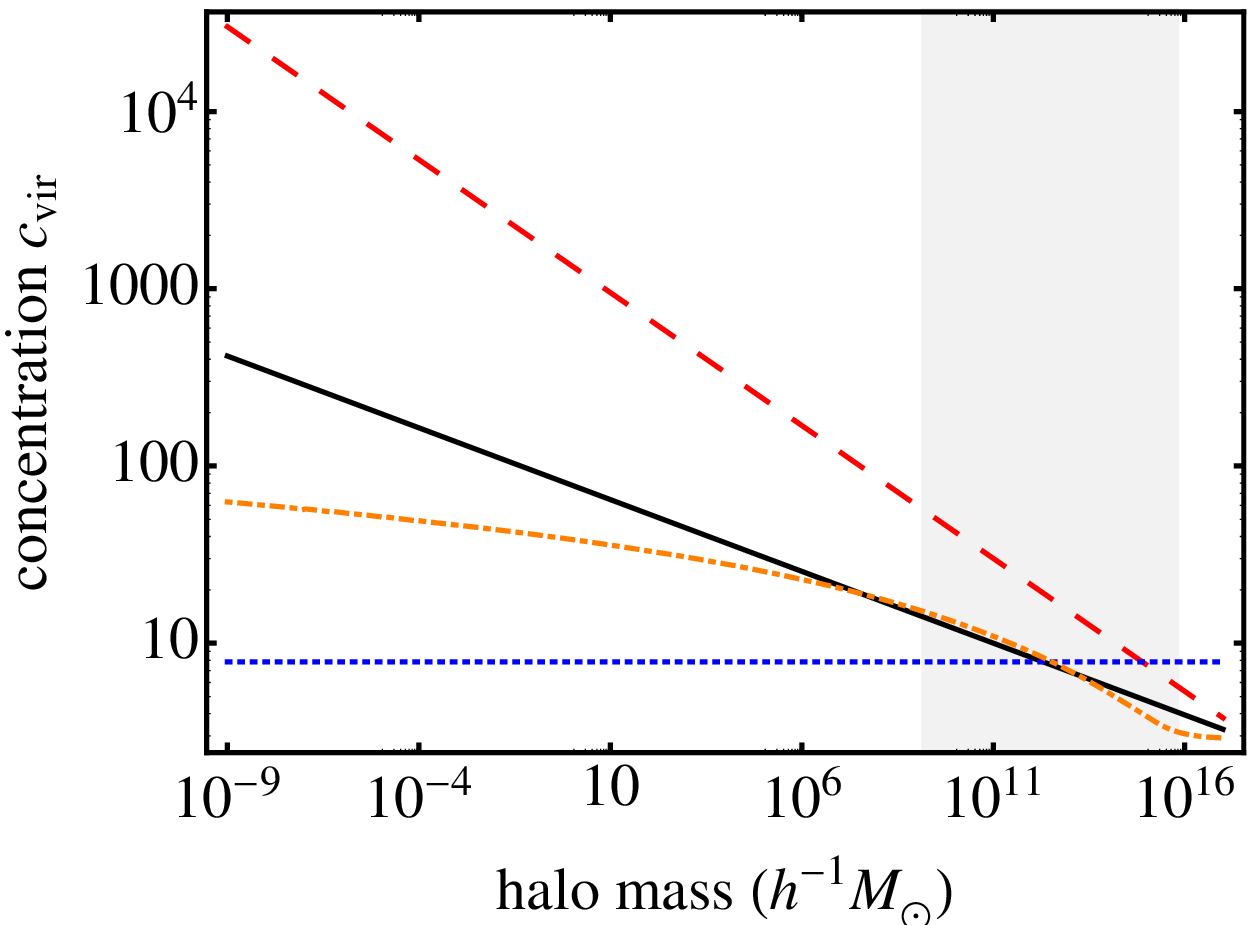}}
\end{center}
\caption{Mass-concentration relations used in this work. The solid (black) line corresponds to the relation in equation (\ref{eq:cD}), from \cite{Duffy2008}; the dashed (red) line corresponds to the relation in equation (\ref{eq:cCN}) from \cite{Comerford2007}; the dotted (blue) line corresponds to the relation in equation (\ref{eq:cflat}) derived from the high-redshift ($z>2$) sample of \cite{Duffy2008}, and the dot-dashed (orange) line corresponds to the relation derived from \cite{Ludlow2013}. The mass range shaded in grey corresponds roughly to the maximum extent where available data make observational comparisons possible; the first three relations have been extrapolated many orders of magnitude below their stated range of validity for the purpose of the calculation in this paper. See \cite{Coe2008}'s Figure 6 for comparison.}
\label{f:concentrationcomparison}
\end{figure}

In Figure \ref{f:concentrationcomparison}, I plot all four mass-concentration relations at redshift 0. In grey, I have shaded in the mass range roughly corresponding to the maximum range where available data make observational comparisons possible.

\section{Results}\label{sec:results}

\begin{figure}[htbp]
\begin{center}
\resizebox{\columnwidth}{!}{\includegraphics{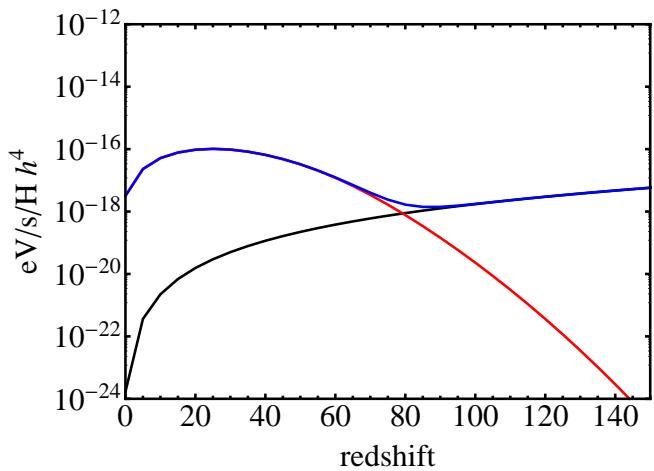}}
\end{center}
\caption{Mean power output per hydrogen nucleus from dark matter annihilation as a function of redshift for the ``fiducial'' model. This model assumes $\langle \sigma v \rangle = 10^{-26}$ cm$^{3}$/s, $m_\chi =$100 GeV, the \cite{Reed2007} halo mass function, a minimum and maximum mass of $M_{\rm{min}}=10^{-9}$ M$_\odot$ and $M_{\rm{max}}=10^{17}$ M$_\odot$, respectively, an NFW halo density profile, and the mass-concentration relation given in equation (\ref{eq:cD}). The black curve is the ``smooth'' component [equation (\ref{eq:sm})], the red curve is the ``structure'' component [equation (\ref{eq:struct})] and the blue curve is the sum of the two components. (Color figures are available in the online version.)}
\label{f:fiducial}
\end{figure}

To illustrate the range of possible predictions for dark matter annihilation power over a range of redshifts, I show, below, the annihilation power calculated for variations on a fiducial model which assumes the following:
\begin{itemize}
\item A dark matter model with $\langle \sigma v \rangle = 10^{-26}$ cm$^{3}$/s and $m_\chi =$100 GeV
\item The halo mass function of \cite{Reed2007}
\item A minimum and maximum mass of $M_{\rm{min}}=10^{-9}$ M$_\odot$ and $M_{\rm{max}}=10^{17}$ M$_\odot$, respectively
\item An NFW halo density profile [equation (\ref{eq:NFW})]
\item The mass-concentration relation in equation (\ref{eq:cD}) from \cite{Duffy2008}
\end{itemize}
This fiducial model is plotted in Figure \ref{f:fiducial}. As can be seen in the plot, the influence of collapsed structures on the annihilation power becomes dominant over the smooth component at late times as structure formation proceeds. The annihilation power from the structure component peaks around the redshift of primordial star formation (roughly between redshifts 10 and 40), at which point it is almost four orders of magnitude stronger than the smooth component. This highlights the importance of understanding both the nature of dark matter clustering and the effects of annihilation on the pre-reionization intergalactic medium. Annihilation occurring in the first protogalaxies might have strong effects on the progress of star formation and reionization, given the pristine state of the gas and the rarity of astrophysical heating sources. I will be explore this possibility further in future work.

The power output scales directly with the quantity $\langle \sigma v \rangle / m_\chi^2$, so the results for other dark matter masses and cross-sections can be straightforwardly derived from the results shown in Figures \ref{f:fiducial}-\ref{f:concentrations}.

\subsection{Influence of mass function} \label{subsec:massfunctionres}

\begin{figure}[htbp]
\begin{center}
\resizebox{\columnwidth}{!}{\includegraphics{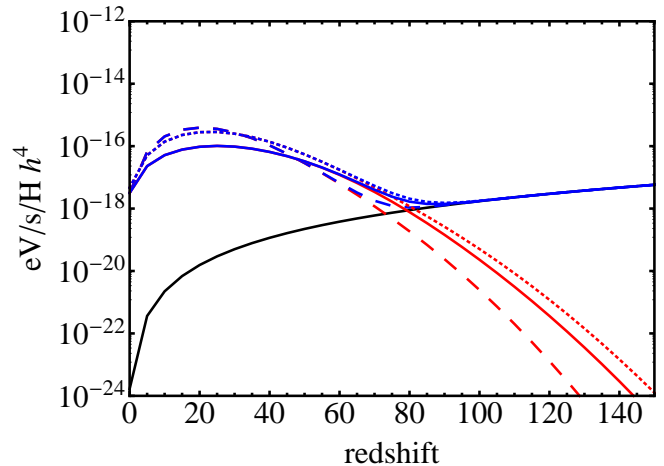}}
\end{center}
\caption{Mean power output per hydrogen nucleus from dark matter annihilation as a function of redshift for the ``fiducial'' model (\cite{Reed2007} halo mass function, solid curves) compared with models using the Press-Schechter mass function (dashed curves) and the Sheth-Tormen mass function (dotted curves). All models assume $\langle \sigma v \rangle = 10^{-26}$ cm$^{3}$/s, $m_\chi =$100 GeV, a minimum and maximum mass of $M_{\rm{min}}=10^{-9}$ M$_\odot$ and $M_{\rm{max}}=10^{17}$ M$_\odot$, respectively, an NFW halo density profile, and the mass-concentration relation given in equation (\ref{eq:cD}). Curve colors are as in Figure \ref{f:fiducial}.}
\label{f:massfunctions}
\end{figure}

In Figure \ref{f:massfunctions}, I show how the choice of halo mass function affects the prediction for the annihilation power as a function of redshift. The largest differences between the mass functions occur at very high redshift ($z \gtrsim 80$) where the smooth component of the annihilation contribution dominates. Even around the redshift of early star formation, however ($z \sim 10-40$), the choice among these standard mass functions can alter the annihilation signal by a factor of 3-4. Alterations in the mass function can also change the redshift at which the structure component begins to dominate over the smooth component, shifting the transition by $\Delta z \approx 10$ between the Press-Schechter and Sheth-Tormen cases.

\subsection{Influence of dark matter halo density profile} \label{subsec:densityprofileres}

\begin{figure}[htbp]
\begin{center}
\resizebox{\columnwidth}{!}{\includegraphics{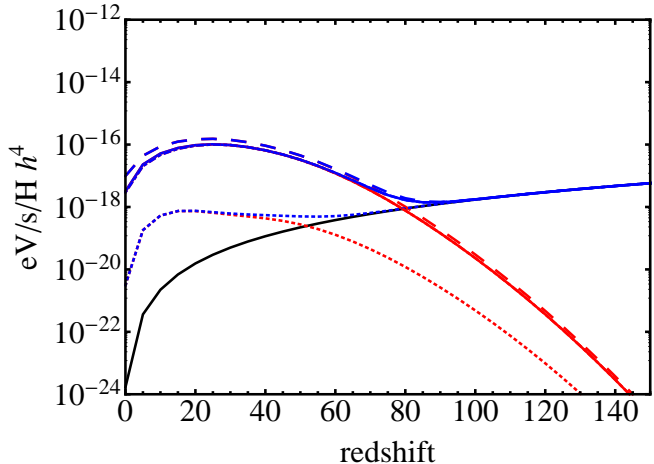}}
\end{center}
\caption{Mean power output per hydrogen nucleus from dark matter annihilation as a function of redshift for the ``fiducial'' model (NFW profile, solid curves) compared to models using the Einasto profile (dashed curves), Burkert profile (dotted curves), and a ``feedback'' model in which halos above 1 \msun{} follow the Burkert model and halos below follow the NFW model (dot-dashed curves). All models assume $\langle \sigma v \rangle = 10^{-26}$ cm$^{3}$/s, $m_\chi =$100 GeV, the \cite{Reed2007} halo mass function, a minimum and maximum mass of $M_{\rm{min}}=10^{-9}$ M$_\odot$ and $M_{\rm{max}}=10^{17}$ M$_\odot$, respectively, and the mass-concentration relation given in equation (\ref{eq:cD}). Curve colors are as in Figure \ref{f:fiducial}.}
\label{f:profiles}
\end{figure}

In this study, I consider three forms for the density profile that have been proposed as fits to the dark matter profiles of galaxies and clusters -- these are plotted in Figure \ref{f:profiles}. For the NFW and Einasto profiles (which are very similar), the exact form of the density profile does not have a strong effect on the annihilation power produced by dark matter halos over most of cosmic history. However, at the lowest redshifts, a difference of up to a factor of 3 can be seen. When all halos have the cored Burkert profile, a dramatic decrease in annihilation power can be seen.

As discussed in Section \ref{subsec:densityprofile}, baryonic feedback effects can also strongly affect the density profiles of dark matter halos, as illustrated by \cite{Pontzen2012,Duffy2010}; and others. To account for the possibility that cores come about only through baryonic feedback, and therefore should not be expected to occur in small halos that do not contain a significant amount of baryonic activity such as star formation, I also plot a case in which halos above 1 \msun{} have the Burkert profile, while halos below this mass have the NFW profile. While this is an arbitrary transition point, it serves to illustrate that feedback effects only acting on massive halos would not significantly alter the dark matter annihilation power. In Figure \ref{f:profiles} below, the ``feedback'' case with cores only in high-mass halos is virtually indistinguishable from the fiducial model. (A very slight difference can be seen at the lowest redshifts in this plot.)

Alterations in the shape of the central density peak coming about through the consideration of warm dark matter models or considerations of alternative formation mechanisms could also significantly alter the annihilation signal. In general, the first would likely decrease the annihilation power by decreasing the central density of the dark matter halos. To determine the effects of alternative formation mechanisms (especially for the smallest halos), more detailed numerical simulations will be needed, and it may be necessary to consider a wider range of density profiles.

\subsection{Influence of minimum mass of dark matter halos} \label{subsec:minmassres}

\begin{figure}[htbp]
\begin{center}
\resizebox{\columnwidth}{!}{\includegraphics{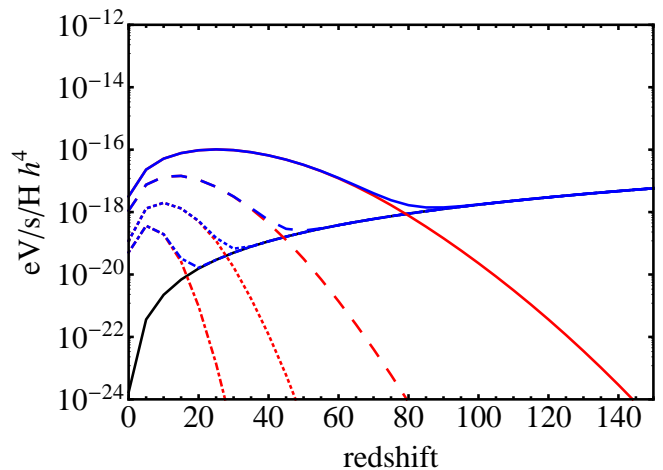}}
\end{center}
\caption{Mean power output per hydrogen nucleus from dark matter annihilation as a function of redshift for the ``fiducial'' cold dark matter model ($M_{\rm{min}}=10^{-9}$, solid curves) compared to a model with a lower mass cut-off of $M_{\rm{min}}=10^{-6}$ M$_\odot$ (dashed curves). Also plotted are warm dark matter cases with lower-mass cut-offs of $10^3$ \msun{} (dotted curves) and $10^7$ \msun{} (dash-dotted curves). All models assume $\langle \sigma v \rangle = 10^{-26}$ cm$^{3}$/s, $m_\chi =$100 GeV, the \cite{Reed2007} halo mass function, an NFW halo density profile, and the mass-concentration relation given in equation (\ref{eq:cD}). Curve colors are as in Figure \ref{f:fiducial}.}
\label{f:masscutoffs}
\end{figure}

Figure \ref{f:masscutoffs} shows the influence of changing the minimum mass of dark matter halos from that of the fiducial cold dark matter model ($10^{-9}$ \msun{}) to an alternative value of $10^{-6}$ \msun{}, as well as WDM cases with cut-offs at $10^3$ \msun{} and $10^7$ \msun{}. As the plot indicates, the lower-mass cut-off of the matter power spectrum has a strong effect on the annihilation power, especially at early times. This is due to the fact that small-mass halos begin forming at high redshift and dominate the halo populations at all times. The strongest differences are at the earliest times; there, the structure component of the annihilation power changes by many orders of magnitude depending on the lower-mass cut-off chosen, though at redshifts above about 80, the smooth component dominates the signal for all models. Even if the signal is dominated by the smooth component, however, this high-redshift difference may be important for the local effect it can have on baryons in the first collapsed halos.

Considering just cold dark matter, the lower-mass cut-off can shift the redshift at which collapsed halos dominate the annihilation power by $\Delta z \sim 35$, from $z \approx 80$ in the fiducial case to $z \approx 45$ in the $M_{\rm{min}}=10^{-6}$ \msun{} case. At $z \approx 35$, when the collapsed component dominates in both cases, the difference between the power output predictions is approximately two orders of magnitude. Even at the lowest redshifts, assuming the larger value for the lower-mass cut-off decreases the expected annihilation power by a factor of at least 3.

When WDM is considered, the differences are much more pronounced. In the most extreme case, where a cut-off occurs at $10^7$ \msun{}, the structure component doesn't dominate the signal until the epoch of star formation has already begun, at redshifts around 20. Of course, with warm dark matter models, the formation of stars and galaxies is also altered from the fiducial case by the delayed collapse of structures. 

\subsection{Influence of mass-concentration relation} \label{subsec:concentrationres}

Mass-concentration relations are important for predictions of the dark matter annihilation power. As the concentration determines the central density of a halo, it is clear that the annihilation power, which depends on the square of the density, should be strongly affected.

\begin{figure}[htbp]
\begin{center}
\resizebox{\columnwidth}{!}{\includegraphics{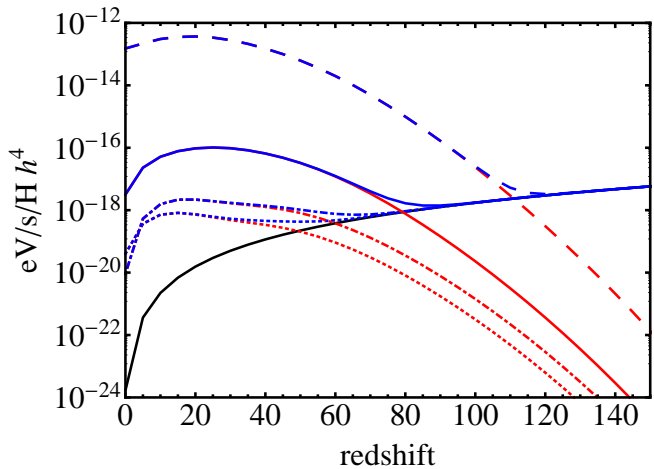}}
\end{center}
\caption{Mean power output per hydrogen nucleus from dark matter annihilation as a function of redshift for the ``fiducial'' model [mass-concentration relation given in equation (\ref{eq:cD}), solid lines] compared to models with the mass-concentration relations given in equation (\ref{eq:cCN}) (dashed curves) and (\ref{eq:cflat}) (dotted curves), and from \cite{Ludlow2013} (dot-dashed curves). All models assume $\langle \sigma v \rangle = 10^{-26}$ cm$^{3}$/s, $m_\chi =$100 GeV, the \cite{Reed2007} halo mass function, a minimum and maximum mass of $M_{\rm{min}}=10^{-9}$ M$_\odot$ and $M_{\rm{max}}=10^{17}$ M$_\odot$, respectively, and an NFW halo density profile. Curve colors are as in Figure \ref{f:fiducial}.}
\label{f:concentrations}
\end{figure}

As can be seen in Figure \ref{f:concentrations}, using the relation from \cite{Comerford2007} [equation (\ref{eq:cCN})] predicts an annihilation power that is between two and four orders of magnitude larger than the fiducial case. This relation, derived from observations of the dark matter in clusters, should not be considered to be the most likely mass-concentration relation for small halos, but it serves to illustrate how important the choice of mass-concentration relations can be for these predictions.

Taking a flat mass-concentration relation from \cite{Duffy2008} shows the limiting case in which the concentration of halos does not evolve at all with mass but only with redshift. Because the smallest halos are the most numerous, removing (or shallowing) the dependence of the mass-concentration relation on mass reduces the annihilation power below the fiducial case. In this example, assuming a flat mass-concentration relation results in a difference of approximately two orders of magnitude in the epoch of early star formation. These alterations in the mass-concentration relation also alter the redshift at which the structure component of the annihilation power begins to dominate, from $z \approx 110$ in the \cite{Comerford2007} case, to $z \approx 80$ in the fiducial case, to $z \approx 50$ in the flat case.

The simulation-derived mass-concentration relation from \cite{Ludlow2013} results in a behavior very similar to that of the flat mass-concentration relation, but with approximately an order of magnitude more power at most redshifts. In this case, the structure component begins to dominate at $z \approx 60$.

\subsection{Range of predictions}\label{subsec:range}

In Figure \ref{f:combined} below, I plot all the models discussed here in one figure to illustrate the full range of predictions. Note that in principle, plausible models could even lie outside this range, as I have not plotted all possible permutations of the halo parameters. This figure thus serves to illustrate a few instances of the broad range of annihilation power predictions possible with varying assumptions about cold dark matter halo properties. Because the WDM cases are qualitatively distinct, I include them as the thin dashed lines that extend outside the shaded region.

\begin{figure}[htbp]
\begin{center}
\resizebox{\columnwidth}{!}{\includegraphics{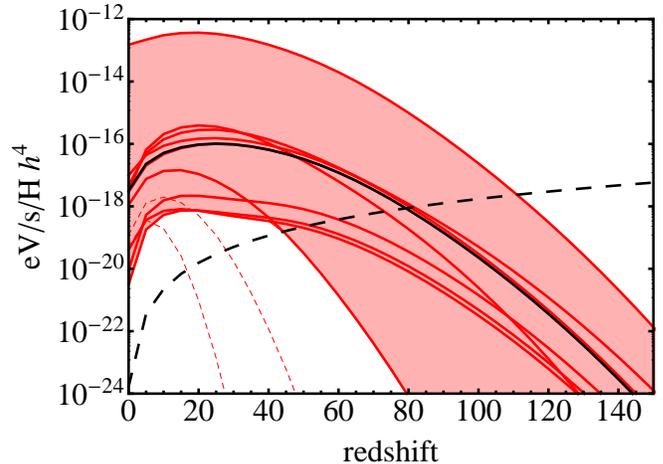}}
\end{center}
\caption{Mean power output per hydrogen nucleus from dark matter annihilation as a function of redshift for the ``fiducial'' model (black solid line) compared to all cold dark matter models shown in Figures \ref{f:massfunctions}-\ref{f:concentrations} above (red solid lines). The black dashed line represents the ``smooth'' component annihilation (which does not depend on the halo model), and the red shading shows the range of annihilation power predictions possible in the cold dark matter models considered here. The WDM cases are included in the thin dashed lines that extend outside the shaded region.}
\label{f:combined}
\end{figure}

\section{Discussion}\label{sec:disc}

\subsection{Halo properties and modelling}\label{subsec:haloprops}

The aim of this work is to demonstrate the range of predictions possible for the dark matter annihilation power history when generic assumptions about dark matter halos are extrapolated to all relevant mass ranges. The form, distribution, evolution and mass scalings of dark matter halos are all integral to a prediction of the annihilation power, especially for the lowest masses. Unfortunately, the lowest masses are the least accessible to observations, necessitating numerical simulations and/or extrapolations across many orders of magnitude in halo mass. As demonstrated by the above results, some of the important remaining questions include:
\begin{enumerate}
\item What form does the dark matter mass function assume?
\item How massive are the smallest collapsed halos, and do they survive to the present day?
\item How do dark matter density profiles evolve with mass and redshift?
\end{enumerate}

These questions, as can be seen in the figures above, have strong effects on our predictions for the dark matter annihilation power, which complicates attempts to determine (or place limits on) the dark matter mass and cross-section from observations. Since an understanding of the annihilation energy input in the evolving intergalactic medium at high redshift can give us insight into the radiation and temperature environment in which the first stars and galaxies were forming, this calculation is a potentially highly consequential one.

\subsection{Velocity dependence of annihilation}\label{subsec:velocity}

Although I have not explored it in this work, the velocity dependence of dark matter annihilation can also affect our predictions for the annihilation power and thus constitutes an important additional uncertainty. It is standard to assume that the quantity $\langle \sigma v \rangle$ is a constant (as is the case for the canonical WIMP), so the relative velocity of dark matter particles does not affect the probability of an annihilation event. However, some theories call for an annihilation rate that is enhanced at either high or low particle velocities \citep[for discussions of consequences and mechanisms of these enhancements, see, e.g.][]{Kuhlen2012,Zavala2011,Galli2009,Kamionkowski2010,Robertson2009,Tulin2013}. One example is the Sommerfeld enhancement \citep{Arkani-Hamed2009,Hisano2004,Lattanzi2009}, sometimes invoked to reconcile dark matter models with cosmic ray experiment signals, which increases the annihilation rate when the dark matter particles are at low velocities. Such an enhancement would decrease the annihilation rate in the very early universe relative to today, allowing a higher cross section to be consistent with both the freeze-out scenario and local hints of possible annihilation signatures. If taken into account here, this enhancement would increase the importance of the ``smooth'' component of the annihilation power relative to that of the ``structure'' component, resulting in implications for the effects of annihilation on gas at high redshift.

Another possibility is that the annihilation rate is enhanced when relative particle velocities are high, as would occur in p-wave annihilation scenarios \citep{Drees1993}. This would have an effect opposite to the Sommerfeld enhancement, in that the annihilation in collapsed (virialized) structures would increase in importance relative to the annihilation in the smooth dark matter background.

Dropping the assumption that $\langle \sigma v \rangle$ is constant is therefore an important possibility to explore in future predictions of dark matter annihilation's affects on the local gas in halos and in the intergalactic medium. A further complication of these issues comes with the consideration of streams and caustics. Caustic structures in dark matter halos can potentially enhance the annihilation signal on small scales \citep{Sikivie1998,Bergstrom2001}, since the $\rho^2$-dependence of the annihilation power results in a higher total annihilation power in strongly inhomogeneous halos than in smooth halos of the same average density. Caustics are notoriously difficult to reliably simulate, though there have been some claims of robust simulations recently \citep[see, e.g.][]{Abel2012}. The phenomenon of caustics highlights another potentially significant small-scale issue in dark matter halos, which is that the inability of cold dark matter to virialize can result in the long-term persistence of streams. Unlike baryonic halos, in which the velocities of particles get well mixed through dissipative particle interactions, dark matter halos can retain particle streams originating from accretion and merger events. This results in a complicated phase-space structure in dark matter halos. Not only does this allow for the formation of caustics, it also creates an inhomogeneity in the velocity structure that can enhance or reduce the annihilation signal if velocity dependences are considered. I will be exploring this scenario in more detail in future work.

\acknowledgements

I would like to thank Stuart Wyithe for discussions that contributed to this work, Aaron Ludlow for fitting functions and discussion of his simulated mass-concentration relations, Alan Duffy for valuable insight into mass-concentration relations in general,  and Nicole Bell, Jeffrey Cooke, Douglas Finkbeiner, Andrew Melatos, Andrew Pontzen, Gregory Poole and Mark Vogelsberger for many helpful discussions. This work was supported by an Australian Research Council Discovery Early Career Researcher Award.

\bibliographystyle{apj_short}
\bibliography{dmcalc}

\end{document}